\documentclass[12pt]{book}
\usepackage{Sussp} 
\usepackage{graphicx}     
\makeindex
\begin{document}
\headings{Galaxies: from Kinematics to Dynamics}
{Galaxies: from Kinematics to Dynamics}
{Michael R Merrifield}
{School of Physics \& Astronomy, University of Nottingham, UK} 

\section{Introduction}

As will be apparent to anyone reading this book, the practitioners of
N-body simulations have an enormous variety of preoccupations.  Some
are essentially pure mathematicians, who view the field as an exciting
application for abstruse theory.  Others enjoy formulating and
tackling mathematically-neat problems, with little concern over
whether the particular restrictions that they impose are also
respected by nature.  Still others are closer to computer scientists,
inspired by the challenge of developing ever more sophisticated
algorithms to tackle the N-body problem, but showing less interest in
the ultimate application of their codes to solving astrophysical
problems.  This contribution is presented from yet another biased
perspective: that of the observational galactic dynamicist.
Observational astronomers tend to use N-body simulations in a rather
cavalier manner, both as a tool for interpreting existing astronomical
data, and as a powerful technique by which new observations can be
motivated.  The aim of this article is to illustrate this profitable
interplay between simulations and observations in the study of galaxy
dynamics, as well as highlighting a few of its shortcomings.

To this end, the text of this chapter is laid out as follows.
Section~\ref{kindynsec} provides an introduction to the sorts of data
that can be obtained in order to study the dynamical properties of
galaxies, and goes on to discuss the intrinsic stellar dynamics that
one is trying to model with these observations, and the role that
N-body simulations can play in this modeling process.
Section~\ref{nbodyhistsec} gives a brief overview of the historical
development of N-body simulations as a tool for studying galaxy
dynamics.  Section~\ref{ellipticalsec} provides some examples of the
interplay between observations and N-body simulations in the study of
elliptical galaxies, while Section~\ref{disksec} provides further
examples from studies of disk galaxies, concentrating on barred
systems.  These sections are in no way intended to be encyclopedic in
scope; rather, by selecting a few examples and examining them in some
detail, the text seeks to give some flavour for the range of what is
possible in this rich field.  Finally, Section~\ref{futuresec}
contains some speculations as to what may lie in the future for this
productive relationship between observations and N-body modeling.

\section{Kinematics and Dynamics}
\label{kindynsec}

The astronomer can glean only limited amounts of information about
galaxies from observations.  Some of these limitations arise from the
practical shortcomings of telescopes, which can only obtain data with
finite signal-to-noise ratios and limited spatial resolution.
However, some of the restrictions are more fundamental -- one can, for
example, only view a galaxy from a single viewpoint, from which it is
not generally possible to reconstruct its full three-dimensional
shape, even if the galaxy is assumed to be axisymmetric (Rybicki
1986).  We must therefore draw a distinction between kinematics, which
are the observable properties relating to the motions of stars in a
galaxy, and dynamics, which fully describe the intrinsic properties of
a galaxy in terms of the motions of its component stars.  Much of the
study of galaxy dynamics involves attempting to interpret the former in
terms of the latter.

Since stars are not the only constituents of galaxies, there is often
additional information that one can glean from other components such
as gas, whose kinematics may be revealed by its emission lines.  These
additional components can also confuse the issue, as selective
obscuration by dust of some regions of a galaxy can have a major
impact on the observed kinematics (e.g.\ Davies 1991).  However, this
text is concerned with N-body models, which are primarily used to
describe the stellar components of galaxies, so here we concentrate
just on the stellar dynamics.  Nevertheless, it should be borne in
mind that no description of a galaxy, particularly a later-type spiral
system, is complete without considering these other components.

\subsection{Kinematics}

We begin by looking at what properties of a dynamical stellar system
are, at least in principle, observable.  The simplest data that one
can obtain is what is detected by an image -- the distribution of light
from the galaxy on the sky, $\mu(x,y)$.  Even for relatively nearby
galaxies, the smallest resolvable spatial element will contain the light
from many stars, so $\mu$ provides a measure of the number of stars
per unit area on the sky.  

\begin{figure}
[htbp]
\centering
\includegraphics[width=12cm,clip,trim=0 0 0 0]{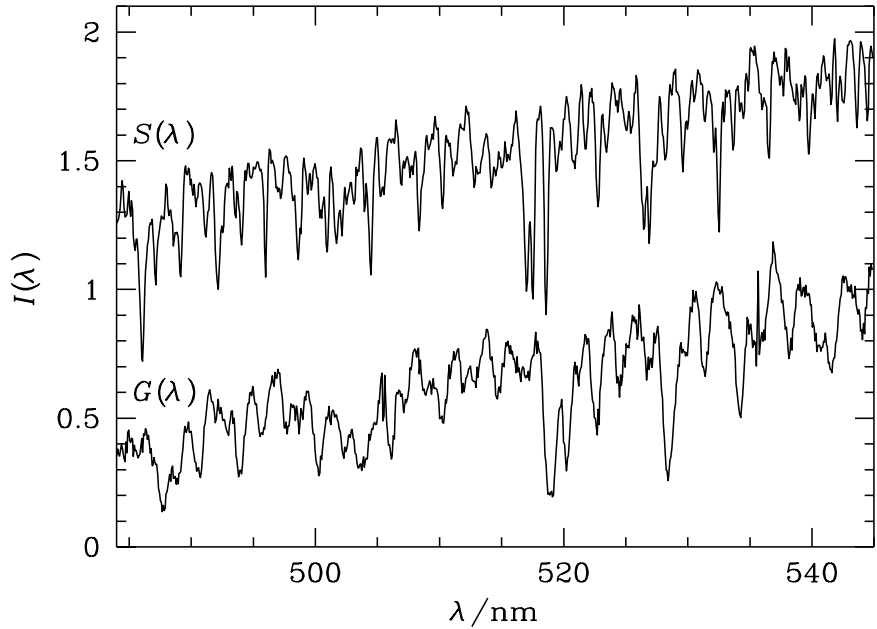}
\caption{Spectra of a star and a galaxy, showing how the absorption
lines in the latter are shifted and broadened relative to the former.}
\label{gsspecfig}
\end{figure}
By obtaining spectra of each of these spatial elements, we can start
measuring the motions of the stars as well as their current locations.
The observed spectrum will be a composite of the light from all the
individual stars.  Stellar spectra contain dark absorption lines due
to the various elements in their atmospheres, but these absorption
lines will be Doppler shifted by different amounts, depending on the
line-of-sight velocities of the stars.  Thus, as
Figure~\ref{gsspecfig} illustrates, the observed absorption lines will
appear broadened and shifted due to the superposition of all the
individual Doppler shifted spectra.  Put mathematically, the observed
spectrum of a galaxy made up from a large number of identical stars will be
\begin{equation}
G(u) = \int {\rm d}v_{\rm los} F(v_{\rm los}) S(u - v_{\rm los}),
\label{kineq}
\end{equation}
where $u = c \ln \lambda$ is the wavelength expressed in logarithmic
units, $S$ is the spectrum of the star in the same units, and
$F(v_{\rm los})$ is the function describing the distribution of stars'
line-of-sight velocities within the element observed.  

Equation~\ref{kineq} is a convolution integral equation, which, in
principle, can be inverted to yield the kinematic quantity of
interest, $F(v_{\rm los})$, for a given galaxy spectrum, $G(u)$, and a
spectrum $S(u)$ obtained using an observation of a suitable nearby
``template'' star.  In practice, such unconstrained deconvolutions are
hopelessly unstable.  The usual approach is therefore to assume some
relatively simple functional form for this function, and adjust its
parameters until Equation~\ref{kineq} is most closely obeyed.  The
best-fitting version of $F(v_{\rm los})$ then provides a model for the
line-of-sight velocity distribution of stars at that point.
Conventionally, and with little physical justification, $F(v_{\rm
los})$ has usually been assumed to be Gaussian, and the fitting
returns optimal values for the mean velocity and dispersion of this
model velocity distribution.  More recently, however, the quality of
data has improved to a point where more general functional forms can
be fitted, allowing a less restricted analysis (e.g.\ Gerhard 1993,
Kuijken \& Merrifield 1993).  With spectra at high signal-to-noise
ratios, it is even possible to attempt a non-parametric analysis,
yielding a best-fit form for $F(v_{\rm los})$ subject only to the most
generic constraints of positivity and smoothness (e.g.\ Merritt 1997).

Although there are many practical difficulties involved in deriving a
completely general description for $F(v_{\rm los})$ [see Binney \&
Merrifield (1998) Chapter 11], it is at least in principle measurable.
Thus, the most general kinematic quantity that one can infer for a
stellar dynamical system is the line-of-sight velocity distribution at
each point on the sky where any of the galaxy's stars are to be found,
$F(x,y,v_{\rm los})$.

\subsection{Dynamics}

To fully specify a galaxy's stellar dynamics, we need to know the
gravitational potential, $\Phi(x,y,z)$, which dictates the motions of
individual stars, and the ``distribution 
function'',\index{distribution function} $f(x,y,z,v_x,v_y,v_z)$, which
specifies the phase density of stars, giving their velocity
distribution and density at each point in the galaxy.

We would therefore appear to have a completely intractable problem,
since we must somehow try to extract the six-dimensional distribution
function from the rather complex observable projection of this quantity,
$F(x,y,v_{\rm los})$, which only has three dimensions.  Fortunately,
however, the form of the distribution function is not completely
arbitrary.  For example, it must be positive or zero everywhere, since
one can never have a negative density of stars.  Further, stars are
(more-or-less) conserved as they orbit around a galaxy, and can only
change their velocities in a continuous manner, dictated by
acceleration due to gravity.  This continuity can be expressed in the
collisionless Boltzmann equation,\index{collisionless Boltzmann equation}
\begin{equation}
{{\rm d} f \over {\rm d}t} = {\partial f \over \partial t} + {\bf
v} \cdot\grad f - \grad\Phi\cdot{\partial f \over \partial{\bf v}} = 0.
\label{boltzmanneq}
\end{equation}

By manipulating the collisionless Boltzmann equation, one can derive a
number of useful formulae for galaxy dynamics.  A full discussion of
this field is beyond the scope of this article, and the interested
reader is referred to the excellent treatment by Binney \& Tremaine
(1987).  Here, we simply summarize some of the key results:
\begin{itemize}
\item By taking a spatial moment of the collisionless Boltzmann
equation, one can derive the virial theorem\index{virial theorem},
which relates the total kinetic and potential energies of the system.
The kinetic energy can be estimated from the observable line-of-sight
motions of stars, from which the potential energy and hence the mass
of the system can be inferred.  It was this approach that provided the
first evidence of dark matter, in clusters of galaxies (Zwicky 1937).
\item By integrating Equation~\ref{boltzmanneq} over velocity, one
obtains the continuity equation\index{continuity equation}, which
describes how the density of stars will vary with time due to any net
flows in their motions.  This equation is central to the dynamics of
``cooler'' stellar systems like disk galaxies, where mean streaming
motions dominate the dynamics; as we shall see below, it has played an
important role in studying the properties of barred galaxies.
\item By multiplying Equation~\ref{boltzmanneq} by powers of velocity
and integrating over velocity, one can derive the Jeans
equations\index{Jeans equations} obeyed by the velocity dispersion,
and their higher-moment analogues.  The Jeans equations describe the random
motions of stars, and have proved particularly important in studies of
the dynamics of elliptical galaxies, where there is little mean
streaming, and random velocities are generally the dominant stellar
motions (e.g.\ Binney \& Mamon 1982).
\item By considering integrals of motion\index{integrals of motion},
one can derive the strong Jeans theorem\index{Jeans theorem}: ``for a
steady state galaxy in which almost all the orbits are regular, the
distribution function depends on at most three integrals of motion.''
For example, in an axi\-symmetric galaxy, one may write
$f(x,y,z,v_x,v_y,v_z) \equiv f(E,J_z,I_3)$, where $E$ is the energy of
the star, $J_z$ its angular momentum about the axis of symmetry, and
$I_3$ is the ``third integral'' respected by the star's orbit, which
cannot generally be written in a simple analytic form.
\end{itemize}
This last result provides us with at least the hope that galaxy
dynamics presents a tractable problem, since we now need only infer a
three-dimensional distribution function from its three-dimensional
observable projection, $F(x,y,v_{\rm los})$.

Equation~\ref{boltzmanneq} describes the continuity equation of a
phase space fluid, which must be solved in order to understand
the dynamics of galaxies.  N-body simulation codes are really just
Monte Carlo integrators tailored to solving this partial differential
equation.  It is very tempting to interpret the bodies in an N-body
code as something more physical, such as the individual stars in a
galaxy.  However, unlike star clusters, galaxies contain so many stars
that current simulations are still several orders of magnitude away
from such a one-to-one correspondence.  It is therefore much healthier
to view an N-body simulation simply as a Monte Carlo solver for the
collisionless Boltzmann equation, which is, in turn, a fluid
approximation to the description of the properties of the large (but
finite) collection of stars that make up a galaxy.

\section{A Brief History of Galaxy N-body Simulations}
\label{nbodyhistsec}

Before launching into a discussion of modern applications of N-body
simulations to studies of galaxy dynamics, it is instructive to look
at the historical development of the field.  N-body simulations of
galaxies date back to well before the invention of the computer.
Probably the first example of the technique was presented by Immanuel
Kant\index{Kant, Immanuel} in his 1755 publication, {\it Universal
Natural History and Theory of the Heavens}.  Part of this book was
concerned with the properties of the Solar System, discussing how the
plane of the ecliptic reflects the ordered motions of the planets
around the Sun, while the more random orbits of comets causes them to
be distributed in a spherical halo.  Kant's N-body simulation involved
using this understanding of the Solar System as an analog computer by
which the Milky Way could be simulated.  He pointed out that the same
law of gravity applies to the stars in the Galaxy as to the planets in
the Solar System.  He therefore argued that the band of the Milky Way
could be understood in the same way as the plane of the ecliptic,
arising from the ordered motion of the stars around the Galaxy.  The
lack of apparent motion in the stars could be explained by the vastly
larger scale of the Milky Way.  He further pointed out that the
scattering of isolated stars and globular clusters far from the
Galactic plane could be compared to comets, their locations reflecting
their more random motions.  Finally, he speculated that other faint
fuzzy nebulae were similar ``island universes'' whose stars followed
similar orbital patterns.  Quite amazingly, Kant's simple analog N-body
simulation had revealed most of the key dynamical properties of
galaxies.

\begin{figure}
[htbp]
\centering
\includegraphics[width=14cm,clip,trim=0 0 0 0]{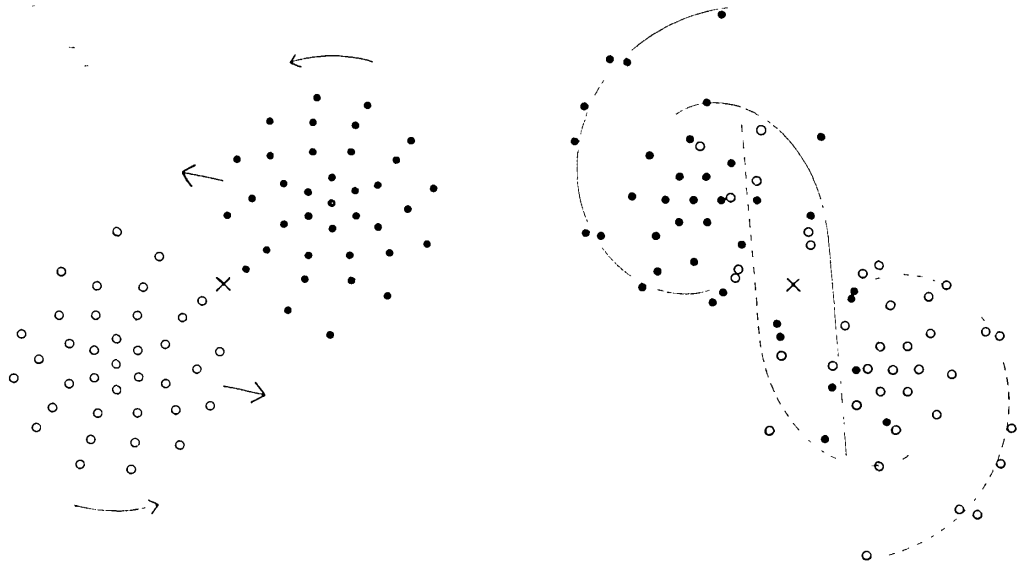}
\caption{Holmberg's original N-body simulation illustrating a merger
between two disk galaxies.  [Reproduced from Holmberg (1941).]}
\label{holmbergfig}
\end{figure}
The next major advance in galaxy N-body simulations was made by
Holmberg\index{Holmberg} (1941).  He used the fact that the intensity
of a light source drops off with distance in the same inverse-square
manner as the force of gravity.  He therefore constructed an analogue
computer by arranging 74 light bulbs on a table: the intensity of light
arriving at the location of each bulb from different directions told
him how large a force should be applied at that position, and hence
how that particular bulb's location should be updated.  With this
analogue integrator, Holmberg was able to show that collisions
between disk galaxies\index{galaxy mergers} can throw off
tidally-induced spiral arms (see Figure~\ref{holmbergfig}), and that
this process can rid the system of sufficient energy that the
remaining stars can become bound into a single object.

\begin{figure}
[htbp]
\centering
\includegraphics[width=14cm,clip,trim=0 0 0 0]{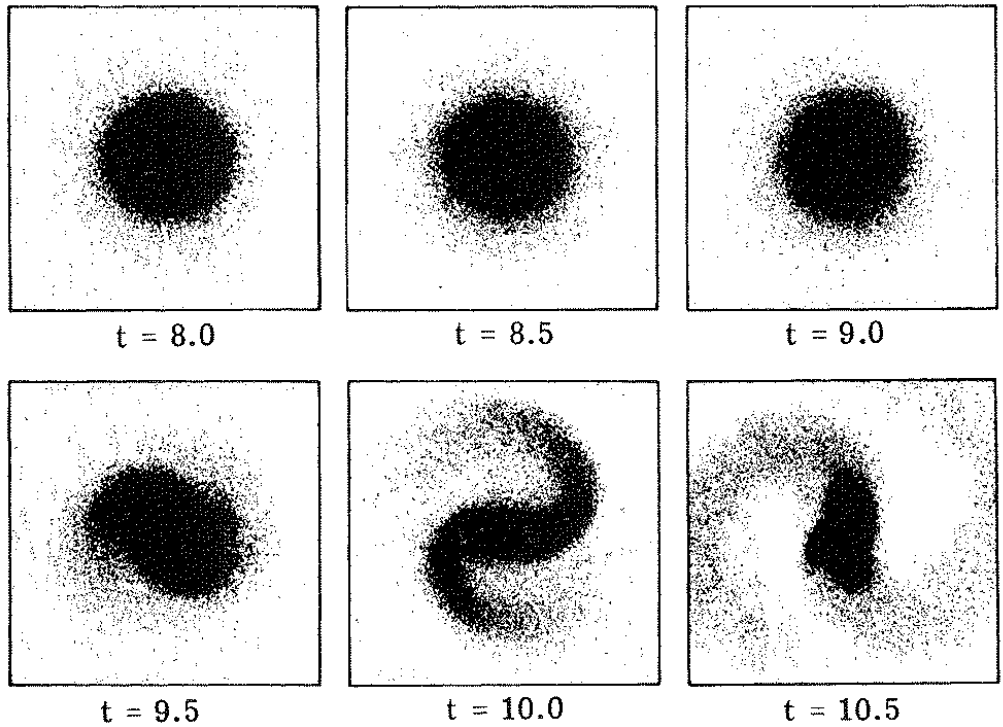}
\caption{N-body simulation of a disk of ``cold'' particles initially
orbiting on orbits very close to circular.  Note the rapid growth of a
strong bar instability. [Reproduced from Hohl (1971).]}
\label{barsimfig}
\end{figure}
The subject really took off in the 1970s with the widespread
availability of digital computers of increasing power.  Numerical
N-body simulations on such a machine allowed Toomre \& Toomre (1972)
to explore the parameter space of galaxy mergers\index{galaxy mergers}
far more thoroughly than Holmberg had been able.  They were thus able
to reproduce the observed morphology of tidal tails and other features
seen in particular merging galaxies, allowing them to reconstruct the
physical parameters of the collisions in these systems.  Other
fundamental insights into galaxies were also made by N-body
simulations around this time, such as the demonstration that a
self-gravitating axisymmetric disk of stars on circular orbits is
grossly unstable, rapidly evolving into a bar\index{barred galaxies}
and spiral arms (see Figure~\ref{barsimfig}).

More recently, progress has been driven by developments in algorithms
and computer hardware, which allow N-body codes to follow the motions
of ever larger numbers of particles.  Although we are still a long way
from being able to follow the motions of the billions of stars that
make up a typical galaxy, the increased number of particles helps
suppress various spurious phenomena that arise from the Poisson
fluctuations in simulations using small numbers of particles.  The
increased number of particles also increases the dynamic range of
scales that one can model within a single simulation.  For example, it
is now possible to look in some detail at the results of mergers
between disk galaxies; it is has long been suggested that such mergers
may produce elliptical galaxies [see Barnes \& Hernquist (1992) for a
review], but the simulations are now so good that we can measure quite
subtle details of the merger remnants' properties such as how fast
they rotate and the exact shapes of their light distributions (Naab
{\it et al.} 1999).  We can then compare these quantities with the
properties of real elliptical galaxies to test the viability of this
formation mechanism.  We are fast reaching the stage where a single
simulation will have sufficient resolution to model simultaneously the
growth of large-scale structure in the Universe and the formation of
individual galaxies (e.g.\ Kay {\it et al.}\ 2000, Navarro \&
Steinmetz 2000).  Thus, within the next few years, we will be able to
perform simulations where the formation and evolution of galaxies can
be viewed within the broader cosmological framework.  However, since
these studies depend critically on the treatment of gas hydrodynamics,
they lie beyond the remit of this article on N-body analysis of the
collisionless Boltzmann equation.

\section{Modeling Elliptical Galaxies}
\label{ellipticalsec}

Elliptical galaxies\index{elliptical galaxies} provide a good place to
start in any attempt to model the stellar dynamics of galaxies.  The
simple elliptical shapes of these systems offers some hope that their
dynamics may also be relatively straightforward to interpret; this
high degree of symmetry means that the assumption of axisymmetry or
even spherical symmetry may not be unreasonable.  Further, the absence
of dust in these systems means that the observed light accurately
reflects the distribution of stars in the galaxy, greatly simplifying
the modeling process.

In fact, elliptical galaxies are so simple that N-body simulations
would not appear to have much of a role to play.  The symmetry of
these systems means that one can readily generate spherical or
axisymmetric models with analytic distribution functions that
reproduce many of the general properties of elliptical galaxies (e.g.\
King 1966, Wilson 1975).  Where one seeks to reproduce the exact
observations of a particular galaxy, Schwarzschild's
method\index{Schwarzschild's method} (Schwarzschild 1979) is often a
much better tool than a full N-body simulation.  This technique
involves adopting a particular form for the gravitational potential --
perhaps, for example, by assuming that the mass distribution follows
the light in the galaxy -- and calculating a large library of possible
stellar orbits in this potential.  One then simply seeks the weighted
superposition of these orbits that best reproduces all the
observational data for the galaxy.  Originally, these fits were made
just to reproduce the projected distribution of stars, but more recent
implementations have also used kinematic constraints such as the
line-of-sight streaming velocities and velocity dispersions at
different projected locations in the galaxy.  It is also possible to
start using information from the detailed shape of the line-of-sight
velocity distribution (e.g.\ Cretton {\it et al.}\ 2000); ultimately,
one could look for the superposition of orbits that reproduces the entire
projected kinematics, $F(x,y,v_{\rm los})$.

\begin{figure}
[htbp]
\centering
\includegraphics[width=14cm,clip,trim=0 0 0 0]{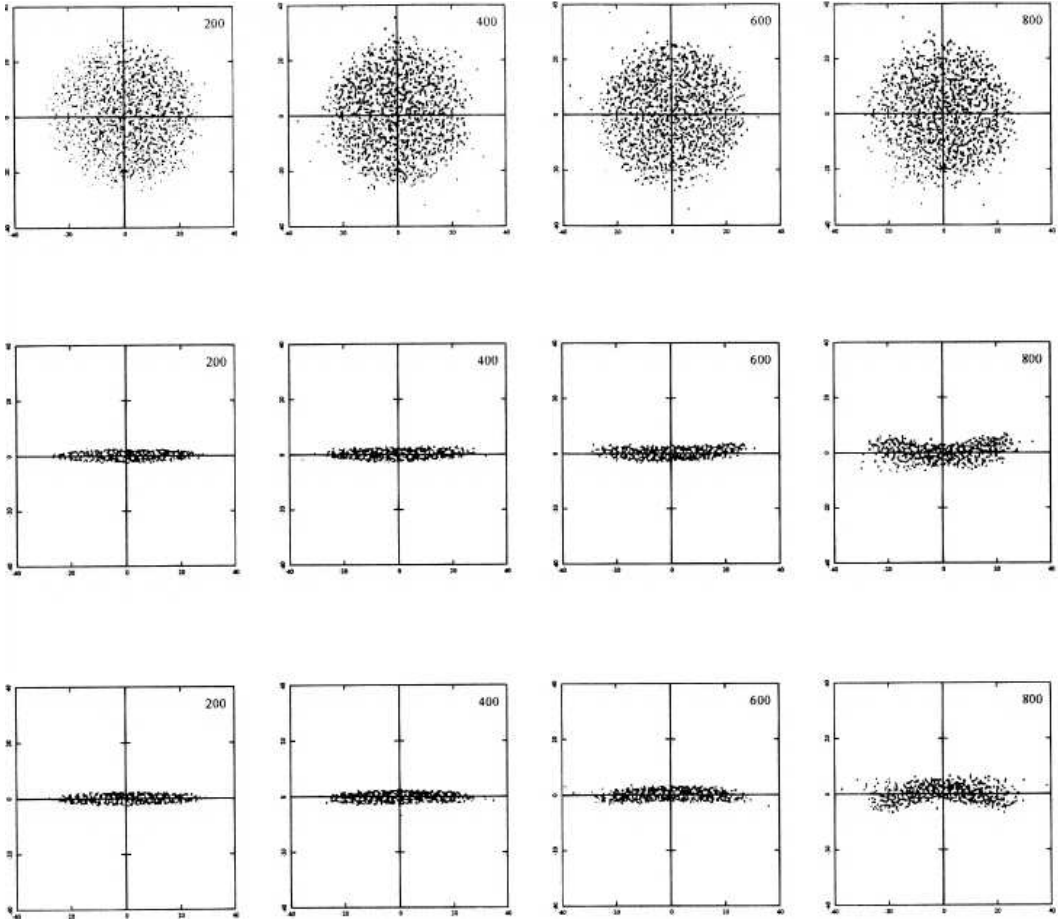}
\caption{N-body simulation of an elliptical galaxy set up in an
initially very flat distribution, as viewed along the three principal
axes.  Note the rapid fattening via a bending instability.
[reproduced from Jessop {\it et al.} (1997).]}
\label{bendingfig}
\end{figure}
There are, however, some aspects of the properties of elliptical
galaxies where N-body simulations offer a powerful tool.  In
particular, if one is concerned with the stability of an elliptical
galaxy, one needs to study the full non-linear time evolution of
Equation~\ref{boltzmanneq}, for which N-body solutions are the most
natural technique.  As an example of the sort of issues one can answer
using this approach, consider the distribution of elliptical galaxy
shapes.  Observations of this distribution have revealed that very
flattened elliptical galaxies do not exist: the most squashed systems
have shortest-to-longest ratios of only $\sim 0.3$.  This observation
could not be explained using the simple modeling techniques described
above, since it is straightforward to derive a distribution function
corresponding to a much flatter elliptical galaxy.  However, if one
takes such a distribution function as the initial solution to
Equation~\ref{boltzmanneq}, and uses an N-body simulation to follow
its evolution, one discovers that it is grossly unstable, usually to
some form of buckling mode, which rapidly causes it to evolve into a
rounder system, comparable to the flattest observed ellipticals (see
Figure~\ref{bendingfig}).  Thus, the absence of flatter elliptical
systems has a simple physical explanation: they are dynamically
unstable.  

Instability analysis using N-body codes has also shed light on other
properties of elliptical galaxies.  For example, Newton \& Binney
(1984) successfully constructed a distribution function that could
reproduce the photometric and kinematic properties of M87: assuming
only that the mass of the galaxy were distributed in the same way as
its light and that the galaxy is spherical, they were able to match
both the light distribution of M87 and the variation in its
line-of-sight velocity dispersion with projected radius.  Thus, they
would appear to have a completely viable dynamical model for M87.
However, Merritt (1987) took this distribution function as the
starting point for an N-body simulation, and showed that the
preponderance of stars on radial orbits at its centre rendered the
model unstable -- the N-body model rapidly formed a bar at its centre.
Thus, the simple spherical model in which the mass followed the light
was invalidated, implying either that M87 is not intrinsically
spherical, or that it contains mass in addition to that contributed by
the stars.

Although some instability analyses can be carried out analytically,
the full calculations of the behaviour of an unstable system,
particularly once the instability has grown beyond the linear regime,
is almost always intractable, making N-body simulations the best
available tool.  Some care must be taken, however, to make sure that
any instability detected is not a spurious effect arising from the
numerical noise in the Monte Carlo N-body integration method (or
indeed, that any real instability is not suppressed by the limitations
of the method).

N-body simulations can also be applied to the study of elliptical
galaxies by providing what might be termed ``pseudo-data.''  When a
new technique is proposed for extracting the intrinsic dynamical
properties of a galaxy from its observable kinematics, one needs some
way of testing the method.  Ideally, one would take a galaxy with
known dynamical properties, and see whether the method is able to
reconstruct those properties.  Unfortunately, it is most unlikely that
the corresponding intrinsic dynamics of a real galaxy would be known
-- if they were, there would be no need to develop the new technique!
However, with an N-body simulation, for which the intrinsic properties
are all measurable, one can readily calculate the appropriate
projections to construct its ``observable'' properties, $F(x,y,v_{\rm
los})$, from any direction.  One can then test the method on these
pseudo-data to see whether the intrinsic properties of the galaxy can
be inferred.

An excellent example of this approach was provided by Statler (1994)
in his attempt to reconstruct the full three-dimensional shapes of
elliptical galaxies from their observable kinematics.  Although these
systems have a simple apparent structure, there is no {\it a priori}
reason to assume that they are axisymmetric, and a more general model
would be to suppose that they are triaxial, with three different
principal axis lengths (like a somewhat deflated rugby ball).  Indeed,
there is strong observational evidence that elliptical galaxies cannot
all be completely axisymmetric.  Images of some ellipticals reveal
that the position angles on the sky of their major axes vary with
radius.  Such ``isophote twist'' cannot occur if a galaxy is
intrinsically axisymmetric, as the observed principal axes of such a
system would always coincide with the projection on the sky of its
axis of symmetry.  Thus, these elliptical galaxies must be triaxial in
structure.  Statler made a study of the dynamics of some simple
triaxial galaxy models, and concluded that one could obtain a much
better measure of the shape of the system by considering the mean
line-of-sight motions of stars as well as their spatial distribution.
As a test of this hypothesis, he took an N-body model, and extracted
from it the observable properties of the mean line-of-sight velocity
and projected density at a number of positions.  Unfortunately, the
constraints on the intrinsic galaxy shape inferred from these data
were found to be only marginally consistent with the true known shape
of the N-body model.  Although in some ways rather disappointing, this
analysis reveals the true power of using N-body simulations to test
such ideas: the N-body simulation did not contain the same simplifying
assumptions as the analytic model that had originally motivated the
proposed idea, so it provided a truly rigorous test of the technique.

As a final example of the way in which N-body simulations can interact
with observations in the study of elliptical galaxies, let us turn to
some work on ``shell galaxies.''\index{shell galaxies} Such systems
typically appear to be fairly normal ellipticals, but careful
processing of deep images reveals that their light distributions also
contain faint ripple-like features in a series of arcs around the
galaxies' centres (e.g.\ Malin \& Carter 1983).  The simplest
explanation for these shells is that they are the remains of a small
galaxy that is merging with the larger elliptical from an almost
radial orbit.  Each shell is made up from stars of equal energy from
the infalling galaxy, which have completed a half-integer set of
oscillations back-and-forth through the larger galaxy, and are in the
process of turning around.  Since the stars slow to a halt as they
turn around, they pile up at these locations, producing the observed
shells.  Shells at different radii contain stars with different
energies, which have completed different numbers of radial orbits
since the merger.  Since the stars in any shell have a very small
velocity dispersion compared to that of the host galaxy, they show up
clearly as sharp edges in the photometry.

\begin{figure}
[htbp]
\centering
\includegraphics[width=14cm,clip,trim=0 0 0 0]{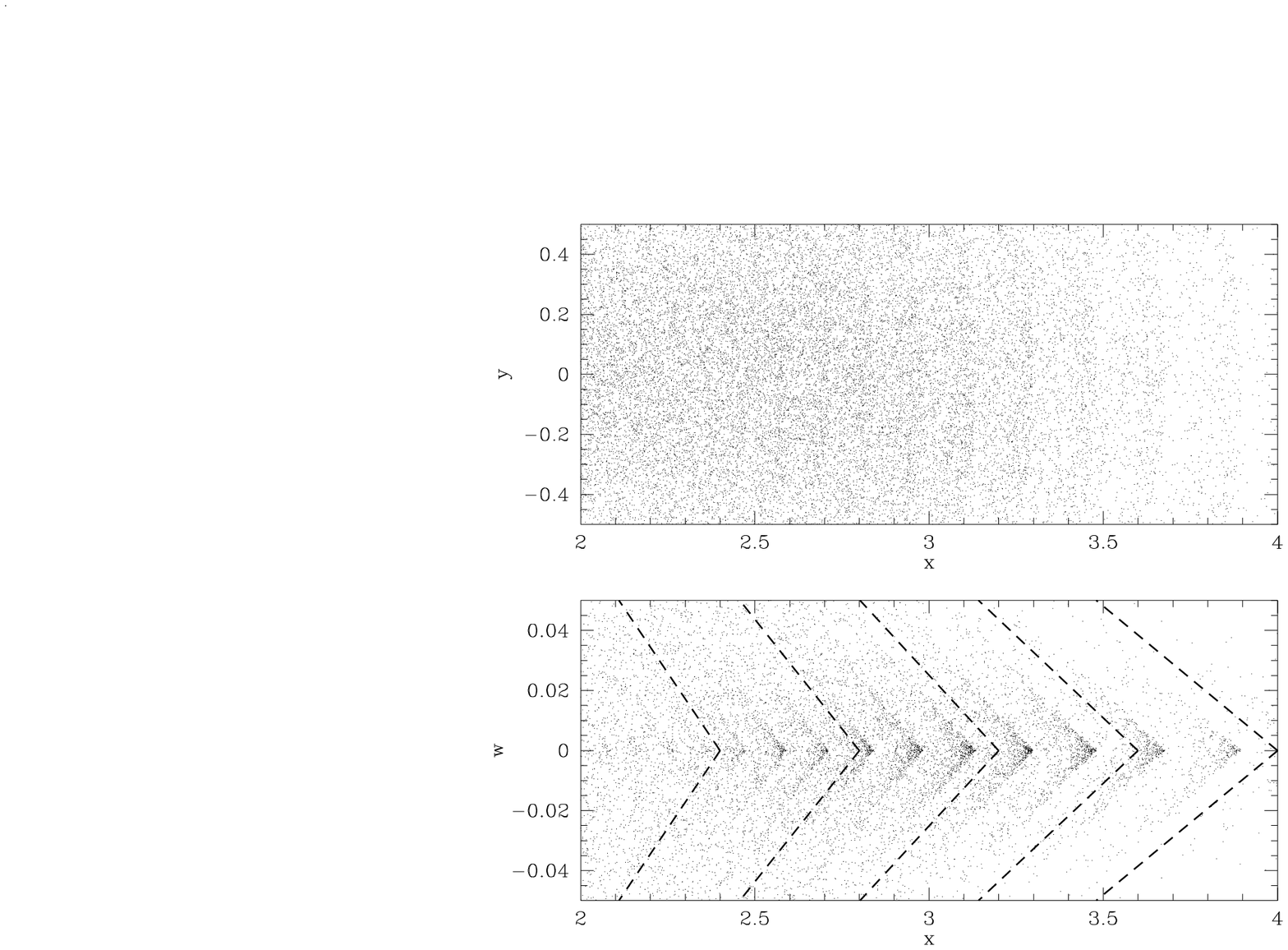}
\caption{N-body simulation, projected to show the observable
properties of the shells created in a minor merger.  The upper panel
shows the photometric properties, while the lower panel shows the
kinematically-observable line-of-sight velocity versus projected
distance along the major axis.  The dashed lines show the predicted
caustic shapes.  [Reproduced from Kuijken \& Merrifield (1998)].}
\label{shellfig}
\end{figure}
N-bodies simulations (e.g.\ Quinn 1984) played a key role in
confirming that such mergers could, indeed, produce sets of faint
shells in the photometric properties of galaxies.  It is therefore
interesting to go on to ask what the most generally-observable
kinematic properties of one of these shells might be.  Again, N-body
simulation offer an excellent tool with which to address this
question.  Figure~\ref{shellfig} presents the results of such a
simulation, showing both the faint photometric shells and the rather
stronger kinematic signature of a minor merger.  The line-of-sight
velocity distribution as a function of position along the major axis
shows a characteristic chevron pattern, whose origins are relatively
straightforward to explain (Kuijken \& Merrifield 1998).  Consider the
stars in a shell whose outer edge lies at $r=r_s$.  By energy
conservation, the radial velocities of stars at $r < r_s$ in this
shell are
\begin{equation}
v_r = \pm \left\{2[\Phi(r_s) - \Phi(r)]\right\}^{1/2},
\end{equation}
where $\Phi(r)$ is the gravitational potential at radius $r$.  By
simple geometry, the observable line-of-sight component of this
velocity is given by
\begin{equation}
v_{\rm los}^2 =\left({z \over r} v_r\right)^2 = 2\left(1 - {x^2 \over
r^2}\right)[\Phi(r_s) - \Phi(r)].
\label{vloseq}
\end{equation}
Close to the shell edge, where $r \sim r_s \ll x$, the maximum value
of $v_{\rm los}$ can be shown, by expanding and differentiating
Equation~\ref{vloseq}, to be
\begin{equation}
v_{\rm max} = \pm \left({1 \over r} {{\rm d}\Phi \over {\rm d} r}
\right)^{1/2} (r_s - x).
\end{equation}
Examples of lines obeying this equation are shown in
Figure~\ref{shellfig}; they clearly match the pattern seen in the
N-body ``observation.''  Thus, if one were to make a detailed kinematic
observation of a shell galaxy and observed this chevron pattern, not
only would one have dynamical evidence for the merger model, but one
would also be able to use the slope of the chevrons to measure ${{\rm
d}\Phi \over {\rm d} r}$ at the radii of each of the shells.
Combining these measurements would allow one to estimate the
gravitational potential of the galaxy in a simple robust manner.

Here, then, is an excellent example of the close interplay that is
possible between observations and N-body simulations.  The photometric
discovery of shells in elliptical galaxies led to a merger theory
that was validated by N-body simulations.  N-body simulations then
provided the motivation for further observations to study the
kinematics of shells in order to make a novel measurement of the
gravitational potentials of elliptical galaxies.  

\section{Modeling Disk Galaxies}
\label{disksec}
We now turn to the use of N-body simulations in the study of disk
galaxies.  Here, the motivation for using N-body modeling is much
clearer.  Spiral galaxies contain a wealth of structure, much of which
is probably transient in nature, so simple analytic models of the type
that do such a good job of describing the basic properties of
elliptical galaxies are clearly inappropriate.  Instead, one needs a
full time-dependent solution to Equation~\ref{boltzmanneq}, for which
N-body simulations provide the most obvious technique.  

It should, however, be borne in mind that the use of
Equation~\ref{boltzmanneq} is often significantly less appropriate in
the study of disk galaxies than was the case for ellipticals.  Active
star formation in many spiral galaxies means that the continuity
implied by the collisionless Boltzmann equation is not strictly valid,
as stars appear in the formation process, and the brightest, most
massive amongst them subsequently disappear in supernovae.  Further,
the location of these star formation regions is largely driven by the
dynamics of the gas from which the stars form.  The collisional nature
of this gas means that it is poorly described by a collisionless
N-body code, and should really be dealt with using much more
sophisticated gas codes.  As a further complication, the dust found in
most spiral galaxies means that a significant fraction of the
starlight is scattered or absorbed.  Thus, there is a rather
complicated relationship between the results of an N-body code (which
essentially gives the distribution of stars in the system) and the
observed photometric properties of a galaxy.  Finally, the likely
transient nature of many of the properties of spiral galaxies also
complicates comparison between observation and theory: since one has
only a snapshot view of a galaxy, one has to search through the
complete evolution in time of an N-body simulation to see if it
matches the observed properties of the galaxy at any point.

Despite these caveats, N-body simulations have provided a wide variety
of insights into the dynamics of disk galaxies.  As for the
ellipticals, N-body simulations have not only been used to explain
many of the observed properties of disk galaxies, but they have also
provided data sets that can test novel analysis techniques, and they
have provided the key motivation for a range of new observations.

As an example of this synergy between N-body simulations and
observations, we consider in some detail the properties of barred
galaxies.\index{barred galaxies} As we have already described in
Section~\ref{nbodyhistsec}, one of the early triumphs of N-body
simulations was in demonstrating that a rectangular bar-like
structure, similar to those seen in more than a third of disk
galaxies, appears due to an instability in a self-gravitating disk of
stars.  Subsequently, as we shall see below, N-body simulations have
enabled us to understand a great deal about the properties of bars.

One of the simplest physical properties of a bar is its pattern speed,
$\Omega_p$, which is the angular rate at which the bar structure
rotates.  In a simulation like that shown in Figure~\ref{barsimfig},
$\Omega_p$ is easy enough to calculate by comparing the bar position
angles at different times.  In a real galaxy, of course, we do not
have the luxury of being able to wait the millions of years required
to see the bar pattern move, so it is less obvious that $\Omega_p$ can
be measured.  However, Tremaine \& Weinberg (1984) elegantly
demonstrated that one can manipulate the continuity equation into a
form that contains only the distribution of stars, their mean
line-of-sight velocities (observable via the Doppler shift in the
starlight at each point in the galaxy), and $\Omega_p$.  Since the
pattern speed is the only unknown, one can derive its value directly
from the other observable properties.  At the time that this technique
was proposed, no observations of barred galaxies had ever produced the
quality of spectral data required to implement the method.  However,
Tremaine \& Weinberg were able to prove its viability by taking a
single snapshot of an N-body simulation and creating
pseudo-observations of the line-of-sight velocities and projected
locations of the objects in it.  The pattern speed derived from this
single pseudo-dataset was found to match that derived from watching
the pattern rotate in the complete time sequence of the simulation.

More recently, kinematic observations have progressed to a point where
this method can be applied to data from real barred galaxies (e.g.\
Merrifield \& Kuijken 1995).  These measurements led to the discovery
that bar patterns seem to rotate rather rapidly, with the bar ends
lying close to the ``co-rotation radius,''\index{co-rotation radius}
which is the radius in the galaxy at which the bar pattern rotates at
the same speed that the stars themselves circulate.  This finding
proved interesting in the light of subsequent N-body simulations of
bars (Debattista \& Sellwood 2000).  These simulations showed that
although bars form with this rapid initial rotation rate, in many
cases the bar pattern speed rapidly decreases almost to a halt.  This
deceleration is the result of dynamical friction:\index{dynamical
friction} the passage of the bar disturbs the orbits of any material
orbiting in the halo of the galaxy, concentrating this material into
``wakes'' of mass that lie behind the rotating bar, exerting a torque
that serves to slow the bar's rotation.  Since cosmological N-body
models of galaxy formation predict that galaxies should form in
centrally-concentrated dark matter halos with plenty of mass at small
radii (e.g.\ Navarro, Frenk \& White 1997), one would expect the
dynamical friction effects from this halo mass to be strong, yielding
slowly-rotating bars.  Thus, either the bars with measured pattern
speeds happen to have been caught very early in their lives when they
have not slowed significantly, or the dark halos in which these barred
galaxies reside do not conform to the cosmologists' predictions.

Finally in this discussion of N-body studies of barred galaxies, let
us turn to the ultimate demise of bars.  Once a bar has grown, there
are several ways that it can be destroyed.  A minor merger with an
in-falling satellite galaxy can put enough random motion into the
stars to mean that they no longer follow highly-ordered bar-unstable
orbits, thus destroying the bar [see, for example, the N-body
simulations by Athanassoula (1996)].  A less violent solution involves
the growth of a massive central black hole in the galaxy.  Inside a
bar, stars shuttle back and forth on ordered orbits aligned with the
bar.  However, N-body simulations have shown that if a central black
hole exceeds a critical mass of a few percent of the bar mass, then
the black hole scatters the passing stars so strongly that they end up
on chaotic orbits that do not align with the bar, thus destroying its
coherent shape (Sellwood \& Moore 1999).  This mechanism is
particularly intriguing, as a bar provides a conduit by which material
can be channeled toward the centre of a galaxy.  If this inflowing
matter is accreted by a central black hole, the central object's mass
can grow to a point where the bar is disrupted, shutting off any
further inflow of material -- a remarkable case of the black hole
biting the hand that feeds it!

\begin{figure}
[htbp]
\centering
\includegraphics[width=8cm,clip,trim=0 0 0 0]{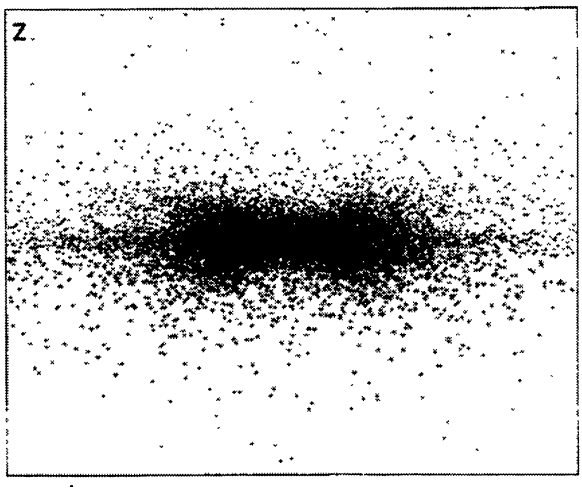}
\caption{N-body simulation showing the peanut-shaped structure
perpendicular to the disk plane into which a bar ultimately evolves.
[Reproduced from Combes \& Sanders (1981)].}
\label{nutfig}
\end{figure}
Even if left in isolation with no mergers or central black holes, thin
bars in disks can have only a very limited lifetime.  N-body
simulations (Combes \& Sanders 1981, Raha {\it et al.}\ 1991) have shown
that bars undergo a buckling instability perpendicular to the plane of
the galaxy, rather similar to that shown in Figure~\ref{bendingfig}.
This instability initially just bends the bar, but the structure then flops
back and forth until it fills a double-lobed fattened region
perpendicular to the galaxy plane, rather like a peanut still in its
shell (see Figure~\ref{nutfig}).

\begin{figure}
[htbp]
\centering
\includegraphics[width=12cm,clip,trim=0 0 0 0]{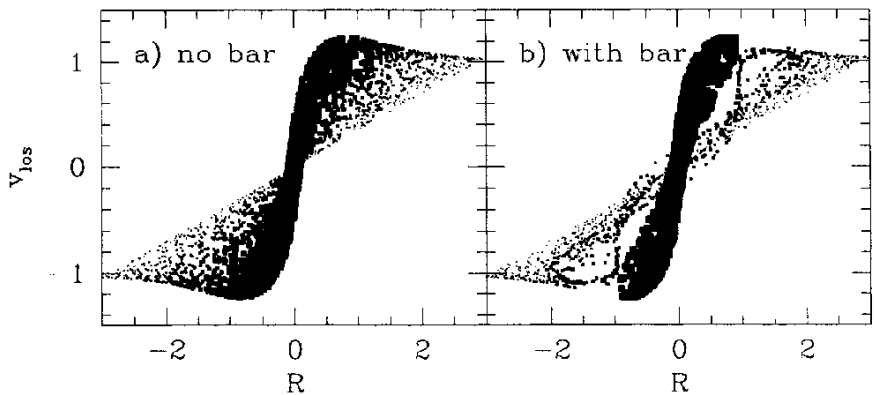}
\caption{Simulations of the observable kinematics (line-of-sight
velocity versus projected radius) along the major axes of edge-on
galaxies, comparing the properties of barred and unbarred systems.
[Reproduced from Kuijken \& Merrifield (1995)].}
\label{barkinfig}
\end{figure}
This N-body discovery has an interesting tie-in with observations: the
bulges of approximately a third of edge-on galaxies are observed to
have boxy or peanut-shaped isophotes, similar to that seen in
Figure~\ref{nutfig} (de Souza \& dos Anjos 1987).  Could it be that
these systems are simply barred galaxies viewed edge-on?  The fraction
certainly corresponds to the percentage of more face-on systems seen
to contain bars, but some more direct evidence is clearly needed.
Again, numerical simulations pointed the way forward: calculations of
orbits in barred potentials have shown that they display a rich array
of structure, with highly elongated orbits, and changes in orientation
at radii where one passes through resonances.  Kuijken \& Merrifield
(1995) investigated the implications of this complexity for the
observable kinematics of edge-on barred galaxies, and showed that the
structure is apparent even in projection: as Figure~\ref{barkinfig}
shows, the changing orientations of the different orbit families shows
up in a rather complex structure in the observable kinematics, $F(x,
v_{\rm los})$.  More sophisticated N-body and hydrodynamic
simulations, allowing for the complex collisional behaviour of
gas, confirm that this structure should also be apparent in the gas
kinematics of an edge-on barred galaxy (Athanassoula \& Bureau 1999).

This N-body analysis motivated detailed kinematic observations of
edge-on disk galaxies, which revealed a remarkably strong correlation:
systems in which the central bulge appears round almost all have the
simple kinematics one would expect for an axisymmetric galaxy, whereas
galaxies with peanut-shaped central bulges almost all display the
complex kinematics characteristic of orbits in a barred potential
(Bureau \& Freeman 1999, Merrifield \& Kuijken 1999).  Thus, the
connection between peanut shaped structures and bars suggested by the
instability found in the N-body models has now been established in
real disk galaxies.  Here, then, is another excellent example of a
case where N-body simulations have not only produced a prediction as
to how galaxies may have evolved to their current structure, but have
also provided the motivation for new kinematic observations that
confirm this prediction.

\section{The Future}
\label{futuresec}

Hopefully, the examples described in this article have given some
sense of the productive interplay between kinematic observations of
galaxies and N-body simulations of these systems, and there is every
reason to believe that this relationship will continue to thrive as
the fields develop.  On the observational side, kinematic data sets
become ever more expansive: the construction of integral field units
for spectrographs has made it possible to obtain spectra for complete
two-dimensional patches on the sky, thus allowing one to map out the
complete observable kinematics of a galaxy, $F(x,y,v_{\rm los})$, in a
single observation.  In the N-body work, developments in computing
power result in ever-larger numbers of particles in the code, allowing
finer structure to be resolved, and giving some confidence that the
results are not compromised by the limitations in the Monte Carlo
solution of Equation~\ref{boltzmanneq}.  More powerful computers also
allow one to analyze the completed N-body simulations more thoroughly:
for example, when comparing transient spiral features in real galaxies
to those in a simulation, one can search through the entire evolution
of the simulation to see whether there are any times at which the data
match the model.

Traditionally, one weakness in combining N-body analysis with
kinematic observations is that although the simulations are very good
at analyzing the generic properties of galaxies, they do not provide a
useful tool for modeling the specific properties of individual
objects.  However, there is now the intriguing possibility that this
shortcoming could be overcome, through Syer \& Tremaine's (1996)
introduction of the idea of a ``made-to-measure'' N-body
simulation.\index{made-to-measure N-body codes} In such N-body
simulations, in addition to its phase-space coordinates, each particle
also has a weight associated with it.  This weight can be equated with
that particle's contribution to the total ``luminosity'' of the model.
Syer \& Tremaine presented an algorithm by which the weights can be
adjusted as the N-body simulation progresses, such that the observable
properties of the model evolve in any way one might wish while still
providing a good approximation to a solution to the collisionless
Boltzmann equation.  Thus, for example, one can take as a set of
initial conditions a simple analytic distribution function, and
``morph'' this model into a close representation of a real galaxy.  In
fact, one can go beyond just the photometric properties of the galaxy,
and match the N-body model to kinematic data as well, thus yielding a
powerful dynamical modeling tool.  Syer \& Tremaine's initial
implementation of this method was fairly rudimentary: for example,
they did not solve self-consistently for the galaxy's gravitational
potential, but instead imposed a fixed mass distribution.  However,
there appears to be no fundamental reason why a more complete
made-to-measure N-body code could not be developed as a sophisticated
technique for modeling real galaxy dynamics.

There has also been a lot of progress in the techniques of stellar
population synthesis (e.g., Bruzual \& Charlot 1993, Worthey
1994).\index{stellar populations synthesis} This approach involves
determining the combination of stellar types, ages and metallicities
that could be responsible for integrated light properties of a galaxy
such as its colours and spectral line strengths.  Thus, one can now go
beyond the simple-minded dynamicist's picture of a galaxy made up from
a large population of identical stars, as assumed in
Section~\ref{kindynsec}; instead, one can begin to pick out the range
of ages and metallicities that could be present in a galaxy, and even
ask whether the different populations have different kinematics.
Here, an extension the made-to-measure N-body approach presents an
exciting possibility.  In addition to a weight, one could associate an
age and a metallicity with each particle.  One could then synthesize
the stellar population associated with that particle, and hence
calculate its contribution to the total spectrum of the galaxy.
Projecting such an N-body model on to the sky, one could calculate the
spectrum associated with any region of the model galaxy by simply
adding up the spectral contributions from the individual particles
(suitably Doppler shifted by their line-of-sight velocities).  By
using the sorts of N-body morphing techniques introduced by Syer \&
Tremaine (1996), one could then evolve an N-body simulation until it
matched the properties of a real galaxy not only in its light
distribution and kinematics, but also in its colours, the strengths of
all its spectral absorption lines, etc.  This complete spectral
modeling -- in essence, a galaxy model that would fit the spatial
coordinates and energy of every detected photon -- would represent the
ultimate match between N-body simulations and observations.  It would
be a truly amazing tool for use in the study of galaxy dynamics, and
would allow us to integrate the evolution of the galaxy's stellar
population into the dynamical picture, opening up a whole new
dimension of information in the study of galaxy formation, evolution
and structure.

\section*{References}

\frenchspacing
\begin{small}

\reference{Athanassoula, E. \& Bureau, M., 1999, ApJ, 522, 699}

\reference{Athanassoula, E., 1996, in Buta R., Croker D.A. and Elmegreen
B.G., eds., Barred Galaxies, Astronomical Society of the Pacific, p.\
307}

\reference{Barnes, J.E. \& Hernquist, L., 1992, ARA\&A, 30, 705}

\reference{Binney, J. \& Mamon, G.A., 1982, MNRAS, 200, 361}

\reference{Binney, J. \& Merrifield, M., 1998, Galactic Astronomy, princeton
University Press}

\reference{Bruzual, A.G. \& Charlot, S., 1993, ApJ, 405, 538}

\reference{Bureau, M. \& Freeman, K.C., 1999, AJ, 118, 126}

\reference{Combes, F. \& Sanders, R.H., 1981, A\&A, 96, 164}

\reference{Cretton, N., Rix, H.-W. \& de Zeeuw, P.T., 2000, ApJ, 536, 319}

\reference{Davies, J., 1991, in Sundelius B., ed, Dynamics of Disc Galaxies,
G\"oteborg, p.\ 65}

\reference{Debattista, V.P. \& Sellwood, J.A., 2000, ApJ, 543, 704}

\reference{de Souza, r.E. \& dos Anjos, S., 1987, A\&A, 70, 465}

\reference{Gerhard, O.E., 1993, MNRAS, 265, 213}

\reference{Hohl, F., 1971, ApJ, 168, 343}

\reference{Holmberg, E., 1941, ApJ, 94, 385}

\reference{Jessop, C.M., Duncan, M.J. \& Levison, H.F., 1997, ApJ, 489, 49}

\reference{Kay, S.T., {\it et al.}, 2000, MNRAS, 316, 374}

\reference{King, I., 1966, AJ, 71, 64}

\reference{Kuijken, K. \& Merrifield, M.R., 1993, MNRAS, 264, 712}

\reference{Kuijken, K. \& Merrifield, M.R., 1995, ApJ, 443, L13}

\reference{Kuijken, K. \& Merrifield, M.R., 1998, MNRAS, 297, 1292}

\reference{Malin, D.F. \& Carter, D., ApJ, 274, 534}

\reference{Merrifield, M.R. \& Kuijken, K., 1995, MNRAS, 274, 933}

\reference{Merrifield, M.R. \& Kuijken, K., 1999, A\&A, 345, L47}

\reference{Merritt, D., 1987, ApJ, 319, 55}

\reference{Merritt, D., 1997, AJ, 114, 228}

\reference{Naab, T., Burkert, A. \& Hernquist, L., 1999, ApJ, 523, L133}

\reference{Navarro, J.F., Frenk, C. \& White, S.D.M., 1997, ApJ, 490, 493}

\reference{Navarro, J.F. \& Steinmetz, M., 2000, ApJ, 538, 477}

\reference{Newton, A.J. \& Binney, J., 1984, MNRAS, 210, 711}

\reference{Quinn, P.J., 1984, ApJ, 279, 596}

\reference{Raha, N., Sellwood, J.A., James, R.A. \& Kahn, F.D., 1991, Nature,
352, 411}

\reference{Rybicki, G.B., 1986, in de Zeeuw P.T., ed, Proc IAU Symp 127, The
Structure and Dynamics of Elliptical Galaxies, Dordrecht, p.\ 397}

\reference{Schwarzschild, M., 1979, ApJ, 232, 236}

\reference{Sellwood, J.A. \& Moore, E.M., 1999, ApJ, 510, 125}

\reference{Statler, T., 1994, ApJ, 425, 500}

\reference{Syer, D. \& Tremaine, S., 1996, MNRAS, 282, 223}

\reference{Toomre, A. \& Toomre, J., 1972, ApJ, 178, 623}

\reference{Wilson, C.P., 1975, AJ, 80, 175}

\reference{Worthey, G., 1994, ApJS, 95, 107}

\reference{Zwicky, F., 1937, ApJ, 86, 217}

\end{small}

\end{document}